\documentclass[11pt]{article}
\usepackage{fullpage,amssymb,amsmath}
\usepackage[dvips]{epsfig}
\newcommand{\nc}{\newcommand} % new command

\nc{\ld}{\left} % left delimiter
\nc{\rd}{\right} % right delimiter

\nc{\ul}[1]{\underline{#1}} % underline once
\nc{\ut}[1]{\underline{\underline{#1}}} % underline twice
\nc{\ct}{\ldots} % to be continued (ellipsis)
\nc{\tr}[1]{\mathrm{#1}} % roman text in mathematical expressions

\nc{\oo}{\infty} % infinity
\nc{\ip}{\cdot} % inner product
\nc{\op}{\times} % outer product
\nc{\nd}{\frac} % fractions
\nc{\ml}{\times} % multiply
\nc{\re}{\textrm{Re}} % real part
\nc{\im}{\textrm{Im}} % imaginary part
\nc{\sr}{\sqrt} % square root
\nc{\su}{\vert} % substitute
\nc{\gt}{\rightarrow} % goes to in limit
\nc{\cv}{\ast} % convolution operator
\nc{\gs}{\ell} % generic subscript or superscript

\nc{\av}[1]{\ld| #1 \rd|} % absolute value
\nc{\iv}[1]{\ld( #1 \rd)} % dependence on independent variables
\nc{\pd}[2]{\partial_{#2} #1} % partial derivative
\nc{\ci}[2]{\ld[ #1, #2 \rd]} % closed interval
\nc{\oi}[2]{\ld( #1, #2 \rd)} % open interval

\nc{\grad}{\nabla} % gradient operator
\nc{\divg}{\nabla \ip} % divergence operator
\nc{\curl}{\nabla \op} % curl operator

\nc{\vt}[1]{\ul{#1}} % vector
\nc{\mr}[1]{\ld[ #1 \rd]} % matrix representation of vectors
\nc{\mc}{\mu\textrm{m}} % micrometer

\nc{\FT}[1]{\tilde{#1}} % Fourier transform
\nc{\DD}{\delta} % Dirac delta function
\nc{\US}{{\cal U}} % unit step function

\nc{\UV}[1]{\vt{u}_{#1}} % unit vector
\nc{\PV}{\vt{r}} % position vector
\nc{\PT}{\iv{\PV, t}} % dependence on position vector and time
\nc{\ZB}[1]{z_{#1}} % material z-axis boundary

\nc{\Co}{c_{0}} % speed of light in vacuum
\nc{\Eo}{\epsilon_{0}} % permittivity of vacuum
\nc{\Mo}{\mu_{0}} % permeability of vacuum
\nc{\No}{\eta_{0}} % impedance of vacuum

\nc{\EF}{E} % macroscopic electric field
\nc{\EI}{E_{i}} % macroscopic incident electric field
\nc{\ER}{E_{r}} % macroscopic reflected electric field
\nc{\BF}{B} % macroscopic magnetic field
\nc{\DF}{D} % macroscopic electric induction field
\nc{\HF}{H} % macroscopic magnetic induction field
\nc{\PF}{P} % polarization field
\nc{\SF}{S} % Poynting vector field
\nc{\Sz}{S_{z}} % axial component of Poynting vector

\nc{\FQ}{\omega} % angular frequency
\nc{\WL}{\lambda_0} % wavelength
\nc{\AF}{\omega} % carrier angular frequency
\nc{\RC}[1]{R_{#1}} % reflection coefficient

\nc{\RP}{\epsilon} % relative permittivity
\nc{\SC}[1]{\chi_{#1}} % susceptibility
\nc{\IR}[1]{n_{#1}} % index of refractive

\nc{\OS}[1]{\alpha_{#1}} % oscillator strength
\nc{\NP}{p_{nl}} % nonlinear parameter
\nc{\AP}[1]{\beta_{#1}} % absorption parameter
\nc{\RW}[1]{\lambda_{#1}} % resonance wavelength
\nc{\RF}[1]{\omega_{#1}} % resonance angular frequency

\nc{\FF}{\vt{F}} % electromagnetic field
\nc{\IF}{\vt{Q}} % induced polarization and magnetization field
\nc{\VD}{\ut{V}} % vacuum propagation dyadic
\nc{\WD}{\ut{W}} % material propagation dyadic
\nc{\EV}{\mr{\EF}} % electric field vector
\nc{\FV}{\mr{\FF}} % electromagnetic field vector
\nc{\IV}{\mr{\IF}} % induction field vector
\nc{\VM}{\mr{\VD}} % vacuum propagation matrix
\nc{\WM}{\mr{\WD}} % material propagation matrix

\nc{\DT}{\Delta t} % time increment
\nc{\DZ}{\Delta z} % space increment
\nc{\SP}{\beta} % stability parameter
\nc{\NS}[1]{N_{#1}} % number of steps

\nc{\PE}{\psi} % pulse envelope
\nc{\TC}{\tau_{0}} % pulse time constant
\nc{\TD}{t_{d}} % pulse time delay
\nc{\UE}{U} % electromagnetic energy density
\nc{\UT}{U_{t}} % total energy density

\nc{\PW}{\vt{\varphi}} % carrier plane wave
\nc{\CF}{\omega_{c}} % carrier angular frequency
\nc{\CP}{\phi} % carrier phase
\nc{\CW}{\lambda_{c}} % carrier wavelength

\def\le{\left(}
\def\ri{\right)}
\def\les{\left[}
\def\ris{\right]}
\def\lec{\left\{}
\def\ric{\right\}}

\def\c#1{\cite{#1}}
\def\l#1{\label{#1}}
\def\r#1{(\ref{#1})}
\def\##1{{\bf #1}}
\def\=#1{\underline{\underline{#1}}}

\begin{document}
\bibliographystyle{unsrt}

\LARGE
\begin{center}
{\bf On the refractive index for a nonmagnetic two--component
medium: resolution of a controversy}

\vspace{10mm} \large

Joseph B. Geddes III\footnote{Corresponding Author. Email:
geddes@uiuc.edu} \\{\em Beckman Institute for Advanced Science and
Technology, University of Illinois at Urbana-Champaign, 405 North
Mathews Avenue, Urbana, IL 61801, USA}\\
\vspace{3mm}
 Tom G. Mackay\footnote{Email: T.Mackay@ed.ac.uk.}\\
{\em School of Mathematics,
University of Edinburgh, Edinburgh EH9 3JZ, UK}\\
\vspace{3mm}
 Akhlesh  Lakhtakia\footnote{Email: akhlesh@psu.edu}\\
 {\em CATMAS~---~Computational \& Theoretical
Materials Sciences Group\\ Department of Engineering Science and
Mechanics\\ Pennsylvania State University, University Park, PA
16802--6812, USA}

\end{center}

\vspace{4mm}

\normalsize

\begin{abstract} The refractive index of a dielectric medium comprising both passive and inverted components in its permittivity was determined using two
methods: (i) in the time domain, a finite--difference algorithm to
compute the frequency--domain reflectance from reflection data for a
pulsed plane wave that is normally incident on a dielectric
half--space, and (ii) in the frequency domain, the deflection of an
obliquely incident Gaussian beam on transmission through a
dielectric slab. The dielectric medium was found to be an active
medium with a negative real part for its refractive index. Thereby,
a recent controversy in the scientific literature was resolved.
\end{abstract}

\noindent {\bf Keywords:} Active medium; Negative refraction;
Time--domain analysis; Frequency--domain analysis

\section{Introduction} \label{S: Introduction}

A complex number $z = \psi\, e^{i\phi}$ possesses two square roots:
 $\sr{z} = \pm\vert\sr{\psi} \vert\,e^{i (\phi/2)}$. This presents a problem when a
 physical quantity (e.g., refractive index) is expressed as the square root of another physical quantity (relative permittivity). Which root is the physical one?

\begin{figure}[h]
\begin{center}
\includegraphics[width = 10 cm]{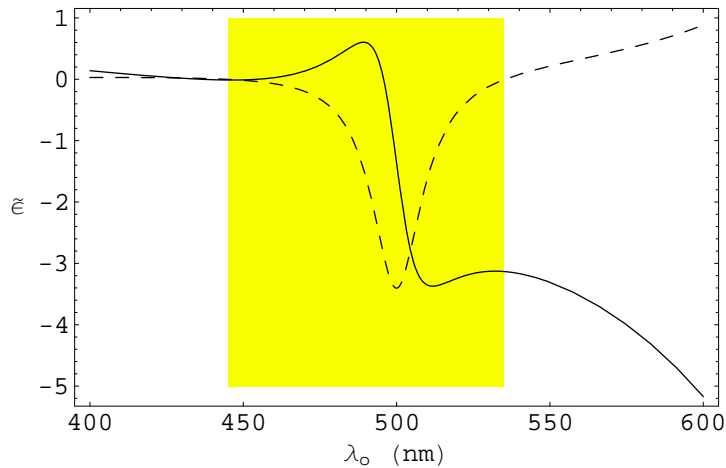}
\caption{Real (solid) and imaginary (dashed) parts of
$\tilde{\epsilon}$ plotted against free--space wavelength (in nm).}
\label{Fig1}
\end{center}
\end{figure}

Our work grows out of a recent controversy over this question in the
context of negative refraction of electromagnetic plane waves. Chen,
Fischer, and Wise (CFW)  considered an isotropic nonmagnetic medium
with a relative permittivity scalar
\begin{equation}
\label{CFW_e} \FT{\RP}\iv{\FQ} = 1 + \sum_{\gs = 1}^{2} \OS{\gs}
\RF{\gs}^{2} \les \RF{\gs}^{2} - (\FQ + i \AP{\gs} \RF{\gs})^{2}
\ris^{-1},
\end{equation}
as a function of the angular frequency $\omega$~\cite{CFW}. Herein,
the constants $\alpha_1 = 2.4401$, $\alpha_2 = -0.14348$, $\beta_1 =
0.028571$,  $\beta_2 = 0.020000$, $\omega_1 =  2.6371$ $\times$ $
10^{15}$ $\mbox{rad} \, \mbox{s}^{-1}$ and $\omega_2 = 3.7673$
$\times$ $10^{15}$ $\mbox{rad} \, \mbox{s}^{-1}$. The permittivity
scalar comprises a passive and an inverted component. Most
importantly, for $\WL\in[445, 535]$~nm, where $\WL$ denotes the
free--space wavelength, ${\rm Im}(\FT{\RP})< 0$, as shown in
Fig.~\ref{Fig1}; here and hereafter, an $\exp(-i\omega t)$
time--dependence is implicit for all frequency--domain field
phasors.

%\begin{table}
%\begin{tabular}{l|l|l|l}
%Parameter & $\gs = 1$ & $\gs = 2$ & Unit\\
%\hline
%$\OS{\gs}$ & 2.4401 & -0.14348 & \\
%$\AP{\gs}$ & 0.028571 & 0.020000 & \\
%$\RF{\gs}$ & 2.6371 $\ml 10^{15}$ & 3.7673 $\ml 10^{15}$ & rad/s
%\end{tabular}
%\caption{Parameters of a dielectric medium considered by Chen,
%Fischer, and Wise~\cite{CFW}.} \label{T: 1}
%\end{table}

CFW deduced that for $\WL\in[445, 535]$~nm, the refractive index
$n=\sqrt{\FT{\RP}}$ must be such that ${\rm Re}(n) < 0$ and ${\rm
Im}(n) > 0$. This deduction stemmed from the premise that the phase
angle of $\FT{\RP}$ is a continuous function of $\WL$, which has
been supported by certain recent theoretical arguments
\cite{Skaar_PRE,Skaar_OL}. Accordingly, CFW argued that their medium
could refract negatively.

Alternatively, on the basis that (i) energy flow grows along the
direction of propagation when  ${\rm Im}(\FT{\RP})<
0$~\cite{Rama_OL}; or (ii) the real part of the wave impedance is
positive--valued on the grounds of causality~\cite{Wei}, it may be
deduced that ${\rm Re}(n) > 0$ and ${\rm Im}(n) <0$ for the CFW
medium  for $\WL\in[445,535]$~nm~\cite{ML_PRL,Rama_PRL}. CFW
disputed this alternative view~\cite{CFW_reply1,CFW_reply2}.

In order to resolve the issue,  two studies were undertaken. First,
following a procedure adopted by Wang and Lakhtakia~\cite{WL_MOTL},
we performed a time--domain calculation that did not explicitly
invoke the frequency--domain concept of refractive index. We used a
finite--difference algorithm to solve the time--domain Maxwell
equations for a pulsed plane wave reflected, at normal incidence,
from a half--space filled with the CFW medium. Then we transformed
the time--domain electric field of the reflected pulse to the
frequency domain and computed the reflectance as a function of
$\WL$. Second, in the frequency domain, we considered a Gaussian
beam  propagating through a slab of the CFW medium at an oblique
angle. The reflection and transmission coefficients were
computed~---~without utilizing the refractive index~---~by solving
the reflection--transmission problem as a boundary--value problem,
and the deflection of the transmitted beam with respect to the
incident beam was determined.

\section{Time--domain analysis} \label{S: Theory}

Let us begin with the time--domain study. Suppose the CFW medium
occupies the half--space $z > \ZB{L}$, ($\ZB{L}>0$), and possesses
the time--domain relative permittivity
\begin{equation}
\RP\iv{t} = \DD\iv{t} +  \sum_{\gs = 1}^{2}\OS{\gs} \RF{\gs}
\exp\iv{- \AP{\gs} \RF{\gs} t} \sin\iv{\RF{\gs} t} \US\iv{t} ,
\end{equation}
where $\DD(t)$ is the Dirac delta function, $\US (t)$ is the unit
step function; note that
\begin{equation}
\FT{\RP}\iv{\FQ} = \int_{-\oo}^{\oo} \RP\iv{t} \exp\iv{i \FQ t} dt
\,.
\end{equation}
The other half--space $z < \ZB{L}$ is vacuous.

A pulsed plane wave propagating in the $+z$ direction is introduced
at $z = 0$, so that for $t > 0$
\begin{equation}
{\bf E} \iv{z=0, t} =\hat{\bf x}\, \ld( \nd{\No \UT}{\TC \sr{\pi}}
\rd)^{\nd{1}{2}} \exp \left[-  \ld( \nd{t - \TD}{\sqrt{2} \TC}
\rd)^{2}\right] \cos \iv{\CF t}.
\end{equation}
The electric field ${\bf E} \iv{z, t}=\hat{\bf x}\,E \iv{z, t}$ is
polarized along the $x$ axis, the magnetic field ${\bf H} \iv{z,
t}=\hat{\bf y}\,H\iv{z, t}$ is polarized along the $y$ axis, $\No =
\sr{\Mo / \Eo}$ is the intrinsic impedance of free space
(permittivity $\Eo$ and permeability $\Mo$), $\Co = 1 / \sr{\Eo
\Mo}$ is the speed of light in free space, $\UT$ sets the energy
density of the pulse, $\TC$ is the time constant, $\TD$ the time
delay, and $\CF$ the carrier frequency.

Upon writing the components of the electromagnetic field in a column
2--vector as $\mr{\FF \iv{z, t}} = \ld[ E\iv{z, t}, H \iv{z, t}
\rd]^{T}$, where the superscript $^{T}$ indicates the transpose, and
substituting the foregoing expressions for $\RP\iv{t}$ into the
Maxwell curl postulates, we found the matrix partial differential
equation %%
\begin{eqnarray} \label{E: MPDE}
\pd {\mr{\,\FF \iv{z, t}}}{t} = \Co \mr{\,\ut{V}\,} \pd{\mr{\,\FF
\iv{z, t}}}{z} - \Eo^{-1}
 \, \pd{\mr{\,\IF\, \iv{z, t}}}{t}
\end{eqnarray}
for $  z, t > 0$. In this equation,
\begin{equation}
\mr{\,\ut{V}\,} = \ld[
\begin{array}{cc}
0 & -\No \\
-1/\No & 0 \\
\end{array} \rd]
\end{equation}
is the vacuum propagation matrix, the column vector
\begin{equation}
\mr{\,\IF \iv{z, t}} = \Eo \int_{0}^{t} \mr{\,\WD \iv{z,t'}}
\mr{\,\FF \iv{z, t - t'}} dt' \, , \label{SAR1}
\end{equation}
and the matrix $\mr{\,\WD \iv{z,t}}$ is null--valued for  $z <
\ZB{L}$, but
\begin{equation}
\mr{\,\WD \iv{z,t}} = \ld[
\begin{array}{cc}
\RP \iv{ t}-\delta(t) & 0 \\
0 & 0
\end{array} \rd].
\end{equation}
for $z > \ZB{L}$. The upper limit on the right side of eqn.
(\ref{SAR1}) accounts for $\mr{\,\FF \iv{z, t }}$ being null--valued
for $t\leq 0$.

We computed the spatiotemporal evolution of the pulsed plane wave
over the domain $\{(z,t)\vert$$ z \in \ci{0}{\ZB{R}},
{\ZB{R}}>{\ZB{L}},t> 0\}$, which was discretized into space steps of
length $\DZ$ and time steps of duration $\DT = \SP \DZ / \Co$, where
$\SP < 1$ is a stability parameter. We discretized eqn. (\ref{E:
MPDE}) and solved it using over the chosen domain using a leapfrog
finite--difference algorithm. Further details of our solution
procedure are reported elsewhere~\cite{GL_JMO}.

\begin{figure}[h]
\begin{center}
\includegraphics[width = 7 cm]{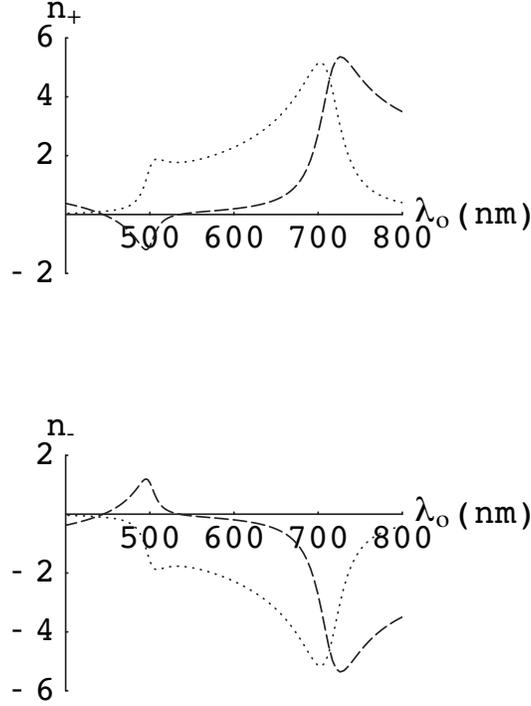}
\caption{Real (dashed) and imaginary (dotted) parts of the indexes
of refraction, as defined in the frequency domain.} \label{F: 1}
\end{center}
\end{figure}

We chose $\ZB{0} = 0$~$\mc$, $\ZB{L}=20$~$\mc$, and $\ZB{R} =
40$~$\mc$. The electric field of the incident pulse $\EI\iv{t}$ was
recorded at $z = 0$~$\mc$, and that of the reflected pulse
$\ER\iv{t}$ at $z = 18$~$\mc$. The incident pulse had the parameters
$\UT = 1$~J~m$^{-1}$, $\TC = 1$~fs, $\TD = 5$~fs, and $\CF = 3.8838
\ml 10^{15}$~$\mbox{rad} \, \mbox{s}^{-1}$, so that its bandwidth
was centered at the free--space wavelength  $\lambda_0 = 485$~nm
with a full--width half--maximum of about $220$~nm. Then, we used
the fast Fourier transform (FFT) to approximate the corresponding
phasors $\FT{\EI}\iv{\WL}$ and $\FT{\ER}\iv{\WL}$ and found the
reflectance from the time--domain calculations as
\begin{equation}
\RC{t}\iv{\WL} = \av{ \FT{\ER}\iv{\WL} \les \FT{\EI}\iv{\WL}
\ris^{-1} }^{2} \,.
\end{equation}

There are two possibilities for the refractive index $n$, {\em
viz.}, $\IR{\pm}$ such that $n_-=-n_+$ and ${\rm Im}(n_+)>0$. Plots
of $\re\iv{\IR{\pm}}$ and $\im\iv{\IR{\pm}}$ as functions of $\WL$
are shown in Fig.~\ref{F: 1}. From these two possibilities, we found
the frequency--domain reflectances
\begin{equation}
\RC{\pm}\iv{\WL} = \av{\les \IR{\pm}\iv{\WL} - 1 \ris \les
\IR{\pm}\iv{\WL} + 1 \ris^{-1}}^{2}\,
\end{equation}
for plane waves normally incident on a half--space. As $\RC{+}
\RC{-} = 1$, $\RC{t}$ can be used to distinguish between them if
$\RC{\pm} \neq 1$~\cite{WL_MOTL}.

\begin{figure}[h]
\begin{center}
\includegraphics[width = 8 cm]{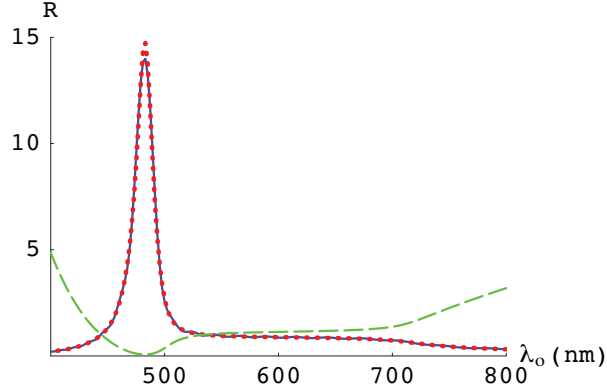}
\caption{Reflectances $R_t$ (blue solid), $R_+$ (red dotted), and
$R_-$ (green dashed).} \label{F: 2}
\end{center}
\end{figure}

Plots of $\RC{t}$, $\RC{+}$, and $\RC{-}$ vs. $\lambda_0$ are shown
in Fig.~\ref{F: 2} over the bandwidth covered by the incident pulse.
As $\vert R_t \vert > 1$ over at least part of that bandwidth, the
CFW medium is active over that part of that bandwidth. However, we
found that neither the pulse reflected from nor the pulse refracted
into the medium grew unboundedly. Furthermore, as the reflectance
$\RC{t}$ obtained from the time--domain calculations closely matches
$\RC{+}$, the refractive index with positive imaginary part (i.e.,
$n_+$) is the correct one.

\section{Frequency--domain analysis} \label{FDA}

\begin{figure}[h]
\begin{center}
\includegraphics[width = 10 cm]{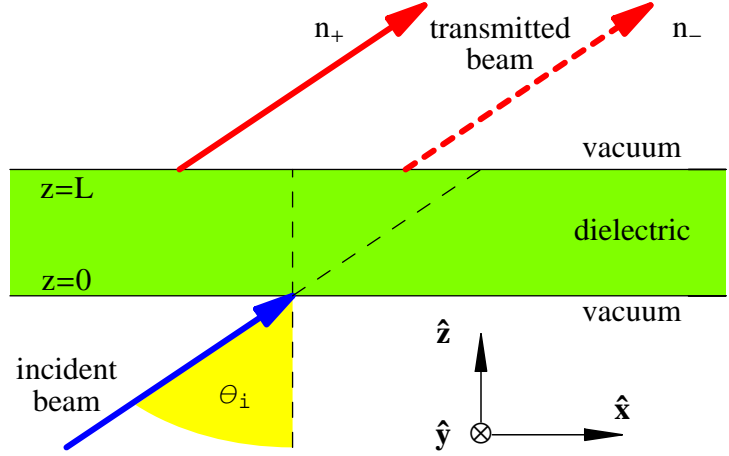}
\caption{A beam is incident onto a slab at a mean angle $\theta_i$
with respect to the unit vector $\hat{\#z}$ normal to the planar
interface. The incident beam strikes the slab at the coordinate
origin (i.e., $x=0, y=0$ and $z=0$). The transmitted beam emerges
from the slab at $z=L$ at a point with  (a) $x> 0$ if the refractive
index of the slab is positive (dashed red arrow); or (b) $x< 0$ if
the refractive index of the slab is negative (solid red arrow).}
\label{schematic_slab}
\end{center}
\end{figure}

Next we turn to the frequency--domain study which does not require
the specification of $n$ as either $n_-$ or $n_+$. Suppose that the
half--space is now replaced by a slab of thickness $L$, as
schematically illustrated in Fig.~\ref{schematic_slab}. The
slab~---~which consists of dielectric material with relative
permittivity defined in eqn. \r{CFW_e}~---~is sandwiched by two
vacuous half--spaces.

A 2D  beam with electric field  phasor \cite{Haus}
\begin{equation}
\tilde{\#E}_i \le x, z, \lambda_0 \ri =
 \int^{\infty}_{-\infty}
  \#e_i (\vartheta) \, \Psi (\vartheta)   \, e^{ i
   \#k_+ (\vartheta)\cdot \#r  }
 \; d \vartheta,  \; \;\; \l{Ei2}
\end{equation}
for $z \leq 0$,  is incident upon the slab at a mean angle
$\theta_i$ relative to the slab normal direction
$\mbox{\boldmath$\hat{z}$}$. The beam is represented as an angular
spectrum of plane waves, with
\begin{eqnarray}
\#k_\pm(\vartheta)& =& k_0 \Big[ \le \vartheta \, \cos \theta_i +
\sqrt{1-\vartheta^2} \, \sin \theta_i \ri \hat{\#x}\nonumber \\ &&
\mp \le \vartheta \, \sin \theta_i - \sqrt{1-\vartheta^2} \, \cos
\theta_i \ri \hat{\#z} \, \Big],
\end{eqnarray}
where $k_0 = 2 \pi / \lambda_0$.
 The angular--spectral function $\Psi (\vartheta)$ is taken to have
 the Gaussian form \c{Haus}
\begin{equation}
\Psi (\vartheta) = \frac{k_0 \, w_0}{\sqrt{2 \pi}} \, \exp \les
-\frac{1}{2} \le k_0 \, w_0 \, \vartheta \ri^2\ris,
\end{equation}
with $w_0$ being the width of the beam waist. Two polarization
states are considered: parallel to the plane of incidence, i.e.,
\begin{eqnarray}
\#e_i (\vartheta) \equiv \#e_\parallel  &=&  \le \vartheta \, \sin
\theta_i - \sqrt{1-\vartheta^2} \, \cos \theta_i \ri
\hat{\#x}\nonumber \\ && + \le \vartheta \, \cos \theta_i +
\sqrt{1-\vartheta^2} \, \sin \theta_i \ri \hat{\#z}
\end{eqnarray}
 and perpendicular to the plane of incidence, i.e.,
\begin{equation}
\#e_i (\vartheta) \equiv \#e_\perp  = \hat{\#y}.
\end{equation}

As the incident beam has the spatial Fourier representation \r{Ei2},
the reflected and the transmitted beams must also have similar
representations. The electric field  phasor of the reflected  beam
is given as
\begin{equation}
\tilde{\#E}_r \le x, z, \lambda_0 \ri = \int^{\infty}_{-\infty}
  \#e_r (\vartheta) \, \Psi (\vartheta)   \, e^{ i
   \#k_-(\vartheta) \cdot \#r
 } \; d \vartheta, \; \;\;  \l{Er}
\end{equation}
for $z \leq 0$, with
\begin{equation}
\#e_r (\vartheta) = \left\{
\begin{array}{c}
r_\parallel \Big[ - \le \vartheta \, \sin \theta_i -
\sqrt{1-\vartheta^2} \, \cos \theta_i \ri \hat{\#x} \\ + \le
\vartheta \, \cos \theta_i + \sqrt{1-\vartheta^2} \, \sin \theta_i
\ri \hat{\#z} \Big] \\ \mbox{for} \;\;\; \#e_i (\vartheta) =
\#e_\parallel \vspace{4pt}
 \\
r_\perp \,\#e_\perp  \;\;\; \mbox{for} \;\;\;   \#e_i (\vartheta) =
\#e_\perp
\end{array}
\right..
\end{equation}
The electric field  phasor of the transmitted  beam is given as
\begin{eqnarray}
\tilde{\#E}_t \le x, z, \lambda_0 \ri = \int^{\infty}_{-\infty}
  \#e_t (\vartheta) \, \Psi (\vartheta)   \, e^{ i
   \#k_+(\vartheta) \cdot \le \#r - L \hat{\#z}   \ri } \; d \vartheta,  \l{Et}
\end{eqnarray}
for $z \geq L$, with
\begin{equation}
\#e_t (\vartheta) = \left\{
\begin{array}{ccr}
t_\parallel \, \#e_i (\vartheta) & \mbox{for} & \#e_i (\vartheta) =
\#e_\parallel \vspace{4pt}
 \\
t_\perp \, \#e_\perp  & \mbox{for} &  \#e_i (\vartheta) = \#e_\perp
\end{array}
\right. .
\end{equation}
The  reflection coefficients $r_{\parallel, \perp}$ and transmission
coefficients $t_{\parallel, \perp}$ were calculated as functions of
$\vartheta$ by solving a boundary--value problem~\cite{STF_book}, as
described in the Appendix.

\begin{figure}[h!]
\begin{center}
\includegraphics[width = 9 cm]{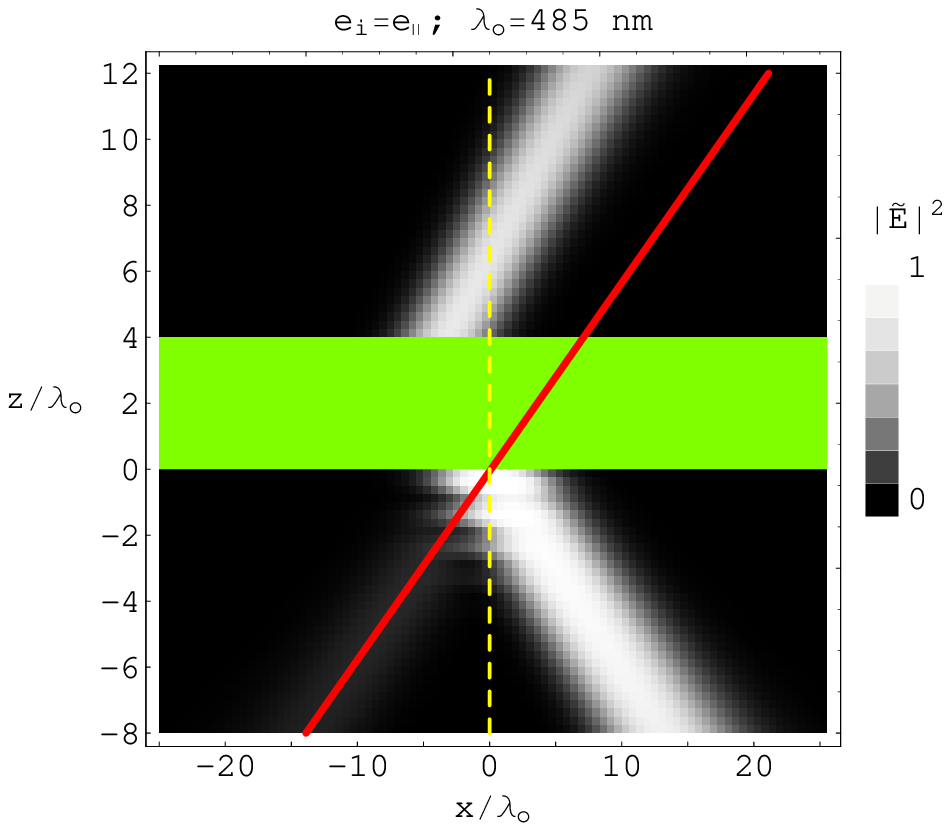}
\includegraphics[width = 9 cm]{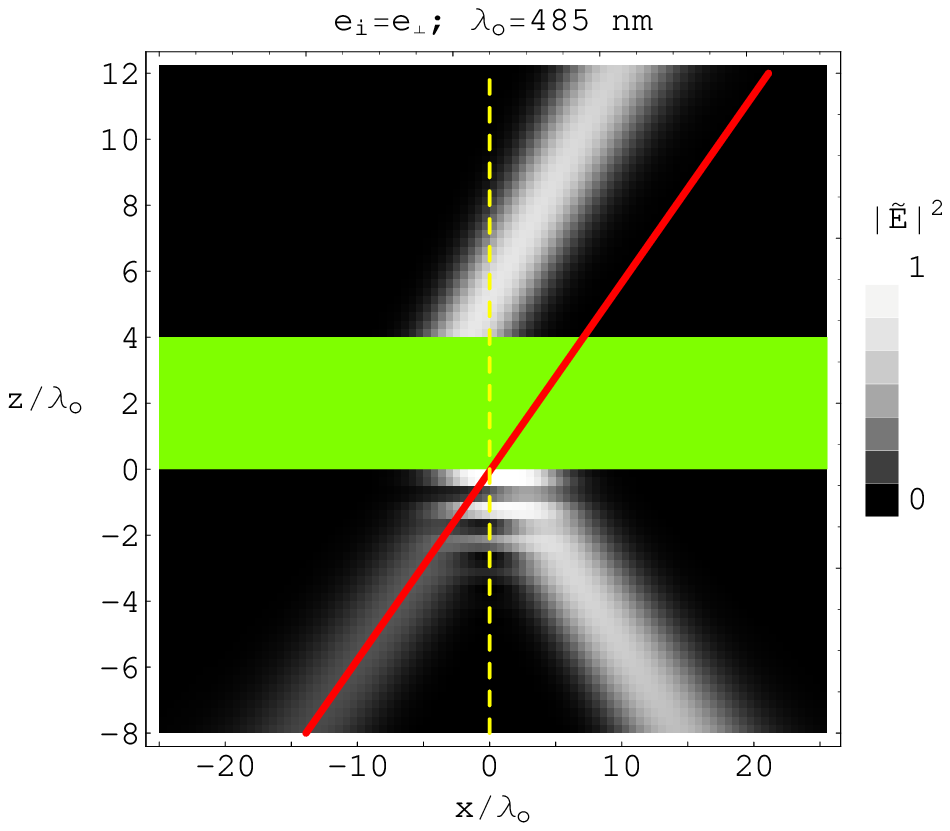}
\caption{Normalized $| \tilde{\#E} |^2$ is mapped in the $xz$ plane
for a 2D Gaussian beam incident onto a CFW dielectric slab at a mean
angle  $\theta_i = 60^\circ$.  $\tilde{\#E}_i$ is polarized parallel
(top) and perpendicular (bottom) to the plane of incidence. The  red
line indicates the mean beam position in the absence of the
dielectric slab.} \label{oblique_beam}
\end{center}
\end{figure}

We fixed the mean angle of incidence $\theta_i = 60^\circ$, the
free--space wavelength $\lambda_0 = 485$ nm, the beam waist $w_0 =
1.75 \lambda_0$, and the slab thickness $L=4\lambda_0$. The
restriction $ \vartheta \in \left[-1,1\right]$ was imposed to
exclude evanescence. The numerical values for the beam waist and
slab thickness were chosen in order to accentuate the clarity of
Fig.~\ref{oblique_beam}, which shows the energy density in both
half--spaces, as defined by
\begin{equation}
 |\tilde{\#E} |^2 = \left\{
 \begin{array}{lcr}
| \tilde{\#E}_i   + \tilde{\#E}_r |^2 & \mbox{for} & z \leq 0 \vspace{4pt} \\
|\tilde{\#E}_t |^2 & \mbox{for} & z \geq L
\end{array}
\right. ,
\end{equation}
for $z / \lambda_0 \in(-8, 12)$ and $ x / \lambda_0 \in (-25, 25)$.

 As illustrated in Fig.~\ref{Fig1}, at
$\lambda_0 = 485$ nm  the relative permittivity of the CFW material
is  $\tilde{\epsilon} = 0.51 - 0.87 i $. The corresponding
reflection and transmission coefficients were numerically determined
at $\vartheta=0$ as $r_\parallel = -3.16 + 0.58 i$, $r_\perp = -1.14
- 1.31 i$, $t_\parallel = \le 3.69 + 3.66 i\ri \times 10^{-8}$ and
$t_\perp = \le -1.82 - 0.10 \ri \times 10^{-8}$. In order to make
visible the tiny fraction of the beam that is transmitted, the
values of $| \tilde{\#E}_t |^2$ in Fig.~\ref{oblique_beam} have been
amplified by a factor of $5 \times 10^{14}$ for $\#e_i =
\#e_\parallel$, and by a factor of $3 \times 10^{15}$ for $\#e_i =
\#e_\perp$. The fact that the CFW medium is active at $\lambda_0 =
485$ nm is clear from $| r_\parallel |^2  > 1$ and  $| r_\perp |^2
> 1$ (in fact, $| r_\parallel |^2 = 10.32 $ and $| r_\perp |^2 = 3.00$).

From Fig.~\ref{oblique_beam} we conclude that the beam undergoes
negative refraction at the two interfaces between the CFW slab and
free space. While this occurs for both  polarization states, it is
more noticeable for the parallel polarization state.

\section{Concluding remarks}

In conclusion, there is an ambiguity inherent in the
frequency--domain concept of refractive index, concerning the choice
of square root. This pertains to both the bending of light at planar
interfaces and the determination of whether the medium under
consideration is active or passive. Using (i) a time--domain method
and (ii) a frequency--domain method, neither of which explicitly
invokes the refractive index, we resolved these issues for the
two--component CFW medium characterized by the relative permittivity
given in eqn. \r{CFW_e}. The CFW medium was found to be (a) an
active medium~---~contrary to the claims of CFW~\c{CFW}; and (b)
negatively refracting~---~contrary to other recent
claims~\cite{ML_PRL,Rama_PRL}.

\vspace{10mm}

\noindent {\bf Acknowledgements:} We thank S.\ A.\ Ramakrishna for
helpful discussions. JBG is supported by a \emph{Beckman
Postdoctoral Fellowship}. TGM is supported by a \emph{Royal Society
of Edinburgh/Scottish Executive Support Research Fellowship}.
\\
\vspace{10mm}

\section*{Appendix}
The reflection coefficients $r_{\parallel, \perp}$ and transmission
coefficients $t_{\parallel, \perp}$ are straightforwardly calculated
by solving  a boundary--value problem as follows~\cite{STF_book}.

Consider the plane wave with electric and magnetic field phasors
\begin{equation}
\left.
\begin{array}{l}
\tilde{\#E}(x,z) = \tilde{\#e}(z,\theta) \, \exp \le i k_0 x \sin
\theta \ri \vspace{4pt}
\\
\tilde{\#H}(x,z) = \tilde{\#h}(z,\theta) \, \exp \le i k_0 x \sin
\theta \ri
\end{array} \right\} \l{eh}
\end{equation}
propagating in the $xz$ plane and incident on  a dielectric slab
with relative permittivity $\tilde{\epsilon}$ occupying the region
between $z= 0$ and $z=L$. The angle $\theta$ is related to the mean
angle $\theta_i$ of a Gaussian beam and the parameter $\vartheta$ by
the twin relations
\begin{equation}
\left.\begin{array}{l} \sin\theta= \vartheta \, \cos \theta_i +
\sqrt{1-\vartheta^2} \, \sin \theta_i \\[5pt]
\cos\theta= -\vartheta \, \sin \theta_i + \sqrt{1-\vartheta^2} \,
\cos \theta_i
\end{array}\right\} \,.
\end{equation}

Substitution of eqn. \r{eh} into the source--free Maxwell curl
postulates $ \nabla \times \tilde{\#E}(x,z) - i \omega \mu_0
\tilde{\#H} (x,z) = \#0$ and $\nabla \times \tilde{\#H}(x,z) + i
\omega\epsilon_0 \tilde\epsilon\tilde{\#E} (x,z) = \#0$ delivers
four differential equations and two algebraic equations. The latter
two equations are easily solved for $\tilde{e}_z$ and $\tilde{h}_z$.
Thereby, the four differential equations may be expressed in  matrix
form as
\begin{equation}
\frac{\partial}{\partial z} \les \, \underline{\tilde{f}} (z,
\theta) \ris = i k_0 \les \,
\underline{\underline{\tilde{P}}}(\theta) \ris \les \,
\underline{\tilde{f}} (z, \theta) \ris, \l{MODE}
\end{equation}
where
\begin{equation}
\les \, \underline{\tilde{f}} (z, \theta) \ris = \les \, \tilde{e}_x
(z, \theta), \,\tilde{e}_y  (z, \theta), \,\tilde{h}_x  (z, \theta),
\,\tilde{h}_y (z, \theta) \ris^T
\end{equation}
is a column vector and
\begin{equation}
\les \, \underline{\underline{\tilde{P}}}(\theta) \,\ris = \les \begin{array}{cccc} 0&0&0& \eta_0 \rho \\
0&0& -\eta_0 &0\\
0& -\tilde{\epsilon} \rho / \eta_0 &0&0\\
\tilde{\epsilon} / \eta_0 & 0 &0&0
\end{array}
\ris
\end{equation}
is a 4$\times$4 matrix with
\begin{equation}
\rho = 1 - \frac{ \sin^2 \theta }{\tilde{\epsilon}}.
\end{equation}
The solution to eqn. \r{MODE} is conveniently expressed as
\begin{equation}
\les \, \underline{\tilde{f}} (L, \theta) \ris = \les \,
\underline{\underline{\tilde{M}}} (L, \theta) \ris \les \,
\underline{\tilde{f}}(0, \theta) \ris, \l{MODE_sol}
\end{equation}
in terms of the transfer matrix~\cite{STF_book}
\begin{equation}
\les \, \underline{\underline{\tilde{M}}} (L, \theta) \ris =
\sum_{\ell=0}^{\infty}\, \frac{1}{\ell!}\,
 \lec i k_0 \les \, \underline{\underline{\tilde{P}}}
(\theta) \ris \, L \ric^\ell\,. \l{Mexp}
\end{equation}
Since the evaluation of $\les \underline{\underline{\tilde{M}}} (L,
\theta) \ris$ as a power series does not invoke the refractive
index, ambiguities associated with the determining the correct
square root of $\tilde\epsilon$ are avoided.

Now we turn to the incident, reflected and transmitted plane waves.
Let the incident plane wave be represented in terms of linear
polarization components as
\begin{equation}
\left.
\begin{array}{r}
\tilde{\#e}_i (z, \theta) = \les a_\perp \, \hat{\#y} +  a_\parallel
\le   \sin \theta \,\hat{\#z} - \cos \theta \,\hat{\#x} \ri \ris \\
\times \exp \le i k_0 z \cos \theta \ri \vspace{4pt}
\\
\tilde{\#h}_i (z, \theta) =  \les a_\perp  \le   \sin \theta
\,\hat{\#z} - \cos \theta \,\hat{\#x} \ri  - a_\parallel \,
\hat{\#y}\,
 \ris \\ \times
\eta^{-1}_0 \, \exp \le i k_0 z \cos \theta \ri
\end{array}
\right\}, z \leq 0.
\end{equation}
The corresponding reflected and transmitted plane waves are given as
\begin{equation}
\left.
\begin{array}{r}
\tilde{\#e}_r (z, \theta) = \les a_\perp r_\perp \, \hat{\#y} +
a_\parallel r_\parallel \le  \cos \theta  \,\hat{\#x} + \sin \theta
\,\hat{\#z} \ri \ris \\ \times \exp \le -i k_0 z \cos \theta \ri
\vspace{4pt}
\\
\tilde{\#h}_r (z, \theta) =  \les a_\perp r_\perp \le \cos \theta
\,\hat{\#x} + \sin \theta \,\hat{\#z} \ri  - a_\parallel r_\parallel
\, \hat{\#y}\,
 \ris \\ \times
\eta^{-1}_0 \,\exp \le -i k_0 z \cos \theta \ri
\end{array}
\right\},  z \leq 0
\end{equation}
and
\begin{equation}
\left.
\begin{array}{r}
\tilde{\#e}_t (z, \theta) = \les a_\perp t_\perp \, \hat{\#y} +
a_\parallel t_\parallel \le   \sin \theta \,\hat{\#z} - \cos \theta
\,\hat{\#x}
\ri \ris \\
\times \exp \les i k_0 (z-L) \cos \theta \ris \vspace{4pt}
\\
\tilde{\#h}_t (z, \theta) =  \les a_\perp t_\perp \le \sin \theta
\,\hat{\#z} - \cos \theta \,\hat{\#x} \ri  - a_\parallel t_\parallel
\, \hat{\#y}\,
 \ris \\ \times
\eta^{-1}_0 \, \exp \les i k_0 (z-L) \cos \theta \ris
\end{array}
\right\},  z \geq L,
\end{equation}
respectively. By application of the boundary conditions at $z=0$ and
$z=L$ to the solution \r{MODE_sol}, the reflection and transmission
coefficients are found to be related by the matrix algebraic
equation
\begin{equation}
  \les \, \underline{\underline{K}} (\theta) \ris \les  t_\perp,  t_\parallel,   0,   0 \ris^T
= \les \,\underline{\underline{\tilde{M}}} (L, \theta) \ris \les \,
\underline{\underline{K}} (\theta) \ris
 \les 1,  1, r_\perp,  r_\parallel  \ris^T,
 \end{equation}
wherein
\begin{equation}
\les \, \underline{\underline{K}} (\theta) \, \ris = \les
\begin{array}{cccc} 0 & - \cos \theta & 0 & \cos
\theta \\
 1 & 0 & 1 & 0 \\
 - \eta_0^{-1} \cos \theta & 0 & \eta_0^{-1} \cos \theta & 0 \\
 0 & - \eta_0^{-1} & 0 & - \eta_0^{-1}
\end{array}
\ris.
\end{equation}
Thus, the reflection and transmission coefficients emerge as
components of the 4$\times$4 matrix
\begin{equation}
\les \, \underline{\underline{\tilde{S}}}\, \ris  =\les \,
\underline{\underline{K}} (\theta) \ris^{-1} \les \,
\underline{\underline{\tilde{M}}} (L, \theta) \ris \les \,
\underline{\underline{K}} (\theta) \ris,
\end{equation}
as per
\begin{eqnarray}
r_\perp &=& -  \les \,\underline{\underline{\tilde{S}}} \,\ris_{31} \le \les \,\underline{\underline{\tilde{S}}} \,\ris_{33}\ri^{-1}, \l{r_perp} \vspace{4pt}\\
r_\parallel &=& -  \les \,\underline{\underline{\tilde{S}}}
\,\ris_{42} \le \les \, \underline{\underline{\tilde{S}}}\,
\ris_{44}\ri^{-1},
 \vspace{4pt}\\
t_\perp &=& \les \,\underline{\underline{\tilde{S}}} \, \ris_{11} -
\les \,\underline{\underline{\tilde{S}}} \,\ris_{13}
\les \, \underline{\underline{\tilde{S}}}\, \ris_{31} \le \les \, \underline{\underline{\tilde{S}}} \,\ris_{33}\ri^{-1} , \vspace{4pt} \\
t_\parallel &=& \les \, \underline{\underline{\tilde{S}}}
\,\ris_{22} - \les \, \underline{\underline{\tilde{S}}}\, \ris_{24}
\les \, \underline{\underline{\tilde{S}}} \,\ris_{42} \le\les \,
\underline{\underline{\tilde{S}}} \,\ris_{44}\ri^{-1} \l{t_perp}.
\end{eqnarray}

\end{document}